\documentstyle [12pt] {article}

\catcode`@=12
\newcommand{\be}{\begin{equation}}
\newcommand{\ee}{\end{equation}}
\newcommand{\ber}{\begin{eqnarray}}
\newcommand{\eer}{\end{eqnarray}}
\newcommand{\bers}{\begin{eqnarray*}}
\newcommand{\eers}{\end{eqnarray*}}
\begin{document}
\vspace{0.5in}
\oddsidemargin -.375in  
\newcount\sectionnumber 
\sectionnumber=0 
\def\be{\begin{equation}} 
\def\ee{\end{equation}}
\thispagestyle{empty}  
\begin{flushright} UTPT-98-09 \\UH-511-852-98\\Fermilab Pub-98/185-T\\
June 1998\
\end{flushright}
\vspace {.5in} 
\begin{center} 
{\Large\bf $\Lambda$ - $\pi$ Phase Shifts in\\  Chiral
Perturbation Theory \\}
\vspace{.5in} 
{{\bf Alakabha Datta{\footnote{email: 
datta@medb.physics.utoronto.ca}}${}^{a,b}$}, {\bf Patrick O'Donnell{\footnote{ 
email: pat@medb.physics.utoronto.ca}}${}^{a}$} and
{\bf Sandip Pakvasa{\footnote{email: pakvasa@uhheph.phys.hawaii.edu}}${}^{c}$} 
\\}
\vspace{.1in} 
${}^{a)}$ {\it Department of Physics and Astronomy,  
University of Toronto, Toronto, Ont., Canada, M5S 1A7.}\\  
${}^{b)}$ {\it Theory Division, Fermilab, P.O. Box 500,   
 Batavia, Illinois, USA, 60510.}\\  
${}^{c)}$ {\it 
Department of Physics and Astronomy, University of Hawaii at Manoa, 
Honolulu, HI 96822, USA.}\\  

\end{center}  

\begin{abstract}
We  calculate  the S and P wave phase  shifts in  $\Lambda - \pi$
scattering at the $\Xi$ mass using the full relativistic $SU(3)_L
\times SU(3)_R$ chiral  perturbation  theory.  We get small phase
shifts similar to previous  calculations  using $ SU(2)_L  \times
SU(2)_R$  chiral  perturbation  theory in the heavy baryon limit.
We also consider possible  off--shell  effects in the coupling of
the Rarita-Schwinger particle $\Sigma(1385)$.  Using $SU(3)$ we
estimate the  off--shell  coupling of the  $\Sigma^*$ to $\Lambda
\pi$ from the  off--shell  coupling  of the  $\Delta$  to $N \pi$
which is obtained  from a fit to the pion--nucleon  data.  We find
that the contributions from the off--shell coupling can be of the
same size as the  other  terms in the  $\Lambda  \pi$  scattering
amplitude.
\end{abstract}  
\newpage
\baselineskip 24pt
\section{Introduction}
The amplitudes for non-leptonic  decays of hyperons are modulated
by the final state  strong  scattering\cite{Fermi}.  These  final
state phase shifts are  necessary in  calculating  the various CP
violating  asymmetries in hyperon decays \cite{pak}.  Some of the
asymmetries  depend  on  $\sin  \delta$  where  $\delta$  is some
combination  of the final  state  scattering  phase  shifts and a
knowledge  of $\delta$ is  necessary to make  predictions  for CP
violations  in  hyperon  decays.  Calculations  of  $\Lambda-\pi$
phase shifts are relevant to the  measurement  of CP violation in
the hyperon decay $\Xi  \rightarrow  \Lambda \pi$ \cite{pak}.  An
experiment  to measure the combined  asymmetry $  \Delta\alpha  =
\Delta\alpha_{\Lambda}  +  \Delta\alpha_{\Xi}  $ is being carried
out   at    Fermilab    \cite{fl}.   Here,   for    example,    $
\Delta\alpha_{\Xi}  = \alpha_{\Xi}  +  \bar{\alpha_{\Xi}}$  where
$\alpha_{\Xi}$   and   $\bar{\alpha_{\Xi}}$   are   the   up-down
asymmetries  in the decay $\Xi \to  \Lambda  \pi$ and its  charge
conjugate process.

The CP violating asymmetry $ \Delta\alpha_{\Xi} $ is proportional
to $  \tan(\delta_S-\delta_P)  $ where  $\delta_S$ and $\delta_P$
are the S and P wave phase shifts in $\Lambda - \pi$  scattering.
There have been  calculations of  $\Lambda-\pi$  scattering phase
shifts in the framework of $SU(2)_L\times  SU(2)_R$ baryon chiral
perturbation theory (HBCHPT) \cite{ml}, with much smaller S wave 
phase shift
than some earlier dispersive  estimates  \cite{Rnath}. The P 
wave phase shift in \cite{ml} was approximately of
the same sign and magnitude as in \cite{Rnath}. The phase
shifts in \cite{ml} were calculated using tree level 
exchanges of low lying
 positive and
negative  parity $\Sigma$  states of spin  $\frac{1}{2}$  and
$\frac{3}{2}$. The fully relativistic calculation of the scattering 
amplitudes includes the higher order $1/m$ corrections to the heavy 
baryon limit. These corrections
should  be naively   suppressed   with  respect  to  the  leading  
 order
contribution   by  factors  of   $\frac{p_\pi}{M_{\Lambda}}$   or
$\frac{M_{\pi}}{M_{\Lambda}}$ where $p_{\pi}$ is the magnitude of
the pion three momentum.  So the subleading  effects are expected
to be at about the 14 \% level . However there may be enhancements 
of these corrections if they are associated with large
coefficients. This is true in the calculations of phase
shifts in the pion-nucleon system where higher order corrections 
in HBCHPT are found to be important for a good 
fit  to  the  data 
\cite{DatPak}.

In this paper we calculate  the S and P wave phase  shifts in the
framework of the fully relativistic  $SU(3)_{L}\times  SU(3)_{R}$
chiral  Lagrangian  and find small phase  shifts for both S and P
waves.  We
also   consider   possible   off--shell   coupling  of  the  spin
$\frac{3}{2}$ $\Sigma (1385)$ resonance, denoted as $\Sigma^*$, to
$\Lambda$  $\pi$.  This off--shell  coupling cannot be determined
from the decay  $\Sigma^*  \to \Lambda \pi$ since it vanishes for
the  on-shell  $\Sigma^*$.  In the SU(3) limit the  magnitude  of
this off--shell coupling can be inferred from the off--shell $\pi
N \Delta$ coupling.

In  Section  2 we  describe  our  formalism  and in  Section 3 we
present our results and conclusions.

\section{ Chiral Lagrangian for Baryons}
The lowest order $SU(3)_{L}\times SU(3)_{R}$ chiral involving the
involving  the $0^-$  mesons,  M and the  $\frac{1}{2}^+$  baryon
octet B can be written as\cite{Georgi}
\ber
L_1 &=& \frac{f^2_{\pi}}{8} Tr(\partial_{\mu}\Sigma \partial^{\mu}
\Sigma^{\dagger})
+iTr({\bar{B}}\gamma^{\mu}\partial_{\mu}B +{\bar{B}}\gamma^{\mu}[V_{\mu},B]) -
mTr({\bar{B}}B)\nonumber\\
&+& D Tr {\bar{B}}\gamma_{\mu}\gamma_5\{A_{\mu},B\} +
F Tr {\bar{B}}\gamma_{\mu}\gamma_5[A_{\mu},B],\
\eer
where
\ber
V_{\mu} &=& \frac{1}{2}
(\xi\partial_{\mu}\xi^{\dagger}+\xi^{\dagger}\partial_{\mu}
\xi) \nonumber\\
A_{\mu} &=& \frac{i}{2}( \xi 
\partial_{\mu}\xi^{\dagger}-\xi^{\dagger}
\partial_{\mu}
\xi) \nonumber\\
\xi &=&\exp(i\frac{M}{f_{\pi}})\nonumber\\
\Sigma &=& \xi^2\
\eer
M and B are the meson and the baryon matrices given by
\bers
M &= &\left [ \begin{array}{ccc}
\frac{\pi^0}{\sqrt{2}}+\frac{\eta }{\sqrt{6}} & {{\pi
 }^+} & {K^+}\\
\pi^- & -\frac {\pi^0}{\sqrt{2}}+\frac{\eta}
{\sqrt{6}} & {K^0}  \\
{K^-} & {\bar{K^0}} & -{\sqrt{\frac{2}{3}}}{{\eta }} \\
\end{array} \right] \\
B &= &\left [ \begin{array}{ccc}
\frac{\Sigma^0}{\sqrt{2}}+\frac{\Lambda }{\sqrt{6}} & {{\Sigma
 }^+} & {p}\\
\Sigma^- & -\frac {\Sigma^0}{\sqrt{2}}+\frac{\Lambda}
{\sqrt{6}} & {n}  \\
{\Xi^-} & {\bar{\Xi^0}} & -{\sqrt{\frac{2}{3}}}{{\Lambda }} \\
\end{array} \right]\
\eers
The transformations of the various fields under  $SU(3)_{L}\times
SU(3)_{R}$ are
\ber
\Sigma  \rightarrow  L\Sigma R^{\dagger}\nonumber\\
\xi  \rightarrow  L \xi U^{\dagger}=U \xi R^{\dagger}\nonumber\\
B \rightarrow UBU^{\dagger}\
\eer
The constants  D$=0.8 \pm 0.14$ and F=$0.5 \pm 0.12$ are obtained
from  a  fit  to  hyperon   semileptonic  decays  \cite{Don}  and
$f_{\pi}=131$ GeV is the pion decay constant.

The spin  $\frac{3}{2}$ $\Sigma (1385)$   belongs  to  the
decuplet  representation  in SU(3).  The interaction  Lagrangian
for the decouplet field $D_{\mu}$ has the general form
\ber
L_{int} & = & g  \bar{D_{\mu}} (g^{\mu\nu} - z \gamma^\mu \gamma^{\nu})
A_{\mu}B + h.c.,\
\eer
where we have  suppressed  the SU(3) indices and we have retained
only terms relevant to our  calculation.  Expanding  $A_{\mu}$ in
terms of the meson fields, the lowest order term  describing  the
interaction of $\Sigma^{\frac{3}{2}+}$(1385) is
\ber
L_{int} & \to & L_2  =  \frac{g}{f_{\pi}}
 \bar{\Sigma^*_{\mu}} [g^{\mu\nu} - z\gamma^\mu \gamma^{\nu}]
\partial_{\nu} \pi \Lambda + h.c.,\
\eer
where  $\Sigma^*_{\mu}$ is the Rarita-Schwinger  field describing
the $\Sigma(1385)$ baryon.  The coupling $g$ may be obtained from
the branching  ratio of $\Sigma^* \to \Lambda \pi$.  The coupling
$z$ remains  undetermined  because  $\gamma_{\mu}u^\mu  =0$ for a
free  Rarita-Schwinger  spinor  $u^{\mu}$  and  so  it  does  not
contribute  to the decay width of  $\Sigma^*  \to  \Lambda  \pi$.
However in the SU(3) limit we can infer this off--shell  coupling
from the $\pi N \Delta$ system.  The $\frac{3}{2}$  propagator is
usually taken as \cite{Pil}
\ber
S_{\mu\nu}^1 & = & \frac{i}{P^2 -M^2}(\gamma
\cdot P +M)[-g_{\mu \nu} +\frac{1}{3}\gamma_{\mu}\gamma_{\nu}
+\frac{2P_{\mu}P_{\nu}}{3M^2} +\frac{\gamma_{\mu}P_{\nu} -
\gamma_{\nu}P_{\mu}}{3M}].\
\eer
On--shell, $S_{\mu \nu}^1$ satisfies the following conditions
\ber
P^{\mu}S_{\mu \nu}^1& = & 0, \nonumber\\
\gamma^{\mu}S_{\mu \nu}^1 & = &0.\
\eer
In the study of the  pion--nucleon  system it has been  suggested
that  the  above  conditions  for  $S_{\mu\nu}$  should  also  be
satisfied off--shell\cite{Joos}.  This leads to a unique form for
the $\frac{3}{2}$ propagator
\ber
S_{\mu\nu}^2 & = & \frac{i}{P^2 -M^2}(\gamma
\cdot P +M)[-g_{\mu \nu} +\frac{1}{3}\gamma_{\mu}\gamma_{\nu}
+\frac{\gamma \cdot P(\gamma_{\mu}P_{\nu}) +
(\gamma_{\nu}P_{\mu})\gamma \cdot P}{3P^2}].\
\eer
The advantage of this form is that the  off--shell  coupling does
not contribute because of the above conditions (Eq.  7) which, in
this case, are also true off--shell and there is no dependence
on the arbitrary  parameter  $z$ in the  amplitude.  In the heavy
baryon limit the  Lagrangian  $L_2$, in standard  HQET  notation,
reduces to
\ber
L_2 & = & \frac{g}{f_\pi} \bar{h^*_{\mu}}
(v) [g^{\mu\nu} - z(v^\mu v^{\nu} -4S^{\mu}S^{\nu})]
\partial_{\nu} \pi h(v) + h.c.,\
\eer
where    $v_{\mu}$    is    the    baryon    four-velocity    and
$S_{\mu}=i\gamma_5\sigma_{\mu\nu}v^{\nu}$  is the spin  operator.
Both the propagators  $S_{\mu\nu}^1$ and $S_{\mu\nu}^2$ reduce in
the heavy baryon limit to
\ber
S_{HB}^{\mu\nu}(\Sigma^*)&
=& -\frac{i}{v \cdot k } 
	\left[ g^{\mu\nu} - v^\mu v^\nu + \frac{4}{3} S^\mu S^\nu \right],\
\eer
where
\bers
P &= & Mv +k
\eers
and the momentum  $k\ll M $  represents  the amount by which P is
off the mass  shell.  In the heavy  baryon  limit the  off--shell
coupling in $L_2$ does not contribute since
$$v_{\mu}S_{HB}^{\mu\nu}=S_{\mu}S_{HB}^{\mu\nu}=0 . $$
So the off--shell  coupling in HBCHPT would  correspond to higher
order   $\frac{1}{M}$   effects.  However  such  effects  can  be
important if these higher order terms are  associated  with large
coefficients and if $z$ is not too small.

The   invariant   transistion   amplitude  for  $\Lambda  -  \pi$
scattering has the general form \cite{Muirhead}
\ber
T & = &\bar{u}_{f} [A(k,\theta) + 
\frac{1}{2}\gamma\cdot(k_1 +k_2)B(k,\theta)]u_i ,\
\eer
where  $\theta$ is the scattering  angle, k is the centre of mass
momentum and $k_1$ and $k_2$ are the incoming and  outgoing  pion
four--momenta.  The scattering amplitude is then given by
\ber
f(\theta)& = & \chi^{\dagger}_f \big [
 f_1 + f_2
\frac{{\bf{\sigma.k_2 \sigma.k_1}}}{k^2}] \chi_i \nonumber\\
&=& \chi^{\dagger}_f \big [
 h + ig
\frac{{\bf{\sigma.(k_2\times k_1)}}}{k^2}] \chi_i \nonumber\\
\eer
with
\ber
f_1 & = &  \frac{E +M}{8 \pi E_{cm}}
\{ A + (E_{cm} -M)B\} \nonumber\\
f_2 & = & \frac{E - M}{8 \pi E_{cm}}\{ -A + (E_{cm} + M)B\} \nonumber\\
h &=& f_1 +f_2 cos\theta \nonumber\\
g &=& f_2 \ 
\eer
where   $\chi_f$,   $\chi_i$   are  the  two   component   spinor
representing the final and intial state $\Lambda$ with mass M and
energy E.  The functions $h$ and $g$  represent the non spin-flip
and the spin-flip amplitude.

The partial waves $f_{L \pm}$ can now be projected out as
\ber
f_{L\pm} & = & \frac{1}{2}\int^{1}_{-1}[P_L(x)f_1 +P_{L \pm 1}(x)f_2]dx\
\eer
where $x=cos\theta$.

Since we are interested in the strong  scattering of the $\Lambda
- \pi$,  system  which is the decay  product in the weak decay of
$\Xi$, the total  angular  momentum of the $\Lambda - \pi$ system
is $J= L \pm  \frac{1}{2}=\frac{1}{2}$  and  hence  the  relevant
partial  waves are $f_{J=0 +  \frac{1}{2}}=f_{0+}$  and $f_{J=1 -
\frac{1}{2}}=f_{1-}$.  The phase  shifts  can then be  calculated
from
\ber
f_{S,P} & = & f_{0+,1-} =\frac{1}{k}e^{i\delta_{S,P}}\sin{\delta_{S,P}}.\
\eer
For small phase shifts we have
\ber
\tan\delta_{S,P}\approx k f_{S,P}.\
\eer
Since we are  calculating  only the tree level  amplitude  we are
unable to satisfy partial wave unitarity.

In our  calculation,  the  $\Lambda-\pi$  scattering  takes place
through the  exchange of $\Sigma$  and  $\Sigma^*(1385)$.  In the
former case both s and u channel amplitudes contribute while only
the u channel contributes for the latter.  The contributions to A
and B for $\Sigma$ exchange are
\ber
A_{\Sigma} & = & (\frac{2D}{\sqrt{6}f_{\pi}})^2
\big[
(M_{\Sigma} +M_{\Lambda})\{2 +(M_{\Sigma}^2-M_{\Lambda}^2)
(\frac{1}{s-M_{\Sigma}^2}+ \frac{1}{u-M_{\Sigma}^2})\}] \nonumber\\
B_{\Sigma} & = & (\frac{2D}{\sqrt{6}f_{\pi}})^2
\big[
(M_{\Sigma} +M_{\Lambda})^2\{
 \frac{1}{u-M_{\Sigma}^2}-\frac{1}{s-M_{\Sigma}^2}\}] \
\eer
where 
$$s=(p_1 +k_1)^2 = (p_2 +k_2)^2 $$
and 
$$ u=(p_2-k_1)^2=(p_1 - k_2)^2$$
with  $p_1,p_2$  being the  initial,  final  baryon  momenta  and
$k_1,k_2$ the initial,  final pion  momenta.  We note that in the
lowest order the F term in the Lagrangian does not contribute.

The contributions  from $\Sigma^*$  exchange using the propagator
$S_{\mu\nu}^1$ are
\ber
A_{\Sigma^*} & = &\frac{g^2}{f_\pi^2}\frac{[
(M +M_\Lambda)
(-M_{\Lambda}^2 +u -3 k_1 \cdot k_2)M^2 + P\cdot k_1
(-M_{\pi}^2-M_{\Lambda}^2 +u)M
+2{(P\cdot k_1)^2}(M + M_{\Lambda}) 
]}{3M^2(M^2-u)} \nonumber\\
& -&\frac{g^2}{f_\pi^2}\frac{2z(u-M_{\Lambda}^2)}{3M} \nonumber\\
& + &
\frac{g^2}{f_\pi^2}z^2\big[\frac{2M_{\Lambda}(u-M_{\Lambda}^2)}{3M^2}
+\frac{4(u-M_{\Lambda}^2)}{3M}] , \nonumber\\
B_{\Sigma^*} & = &\frac{g^2}{f_\pi^2}\frac{[
(-M_{\pi}^2+2M_{\Lambda}^2 +3 k_1 \cdot k_2 +2MM_{\Lambda})M^2 
+ 2P \cdot k_1 M_{\Lambda}M
-2{(P \cdot k_1)^2} 
]}{3M^2(M^2-u)} \nonumber\\
& +& \frac{g^2}{f_\pi^2}z \big[
\frac{4P \cdot k_1}{3M^2} -\frac{4M_{\Lambda}}{3M}] \nonumber\\ 
& + &
\frac{g^2}{f_\pi^2}z^2 \big[-\frac{2M_{\pi}^2}{3M^2}+
\frac{4M_{\Lambda}^2}{3M^2}
-\frac{4P \cdot k_1}{3M^2} +\frac{8M_{\Lambda}}{3M}
] .
\eer
We note that the  off--shell  terms do not have a pole  since the
numerator   in  the   amplitude  of  the   off--shell   terms  is
proportional  to $ (P^2-M^2) $ which cancels the pole term in the
denominator.

If the propagator $S_{\mu\nu}^2$ is used then we have
\ber
A_{\Sigma^*} & = &
\frac{g^2}{f_\pi^2}\frac{
( M + M_{\Lambda})
\{u(-M_{\Lambda}^2 +u - 3 k_1 \cdot k_2) -2P \cdot k_1 M_{\pi}^2\}
}
{3(M^2 -u)u} \nonumber\\
B_{\Sigma^*} & = & \frac{g^2}{f_\pi^2} \frac{
-2 (P \cdot k_1)^2 -u M_{\pi}^2 +2(u + P \cdot k_1)(M_{\Lambda}^2 +M 
M_{\Lambda}) 
+3 u k_1 \cdot k_2
}
{3(M^2 -u)u}\
\eer
As noted  before,  there is no $z$  dependence in this case since
the off--shell terms do not contribute.

\section{Results and Discussions}
In this section we present and discuss our results.  The relevant
masses   and  widths   for  our   calculation   are  taken   from
Ref\cite{PDG}.  The phase shifts from the $\Sigma$  exchange  are
$\delta_S=$  -0.13  degrees and  $\delta_P  =$ -2.84  degrees for
D=0.8.  If we vary D between its limits  0.66-0.94 then we obtain
$\delta_S=$-0.09  degrees to -0.18  degrees  and  $\delta_P=$-1.9
degrees  to  -3.9  degrees.  For  the   $\Sigma^*$   exchange  we
determine  the  coupling  constant  $g$  from  the  width  of the
$\Sigma^* \to \Lambda \pi$,
\bers
\frac{g^2}{f_\pi^2} & = &
\frac{12 \pi \Gamma}{(\frac{M_{\Sigma^*}}{M_{\Lambda}})(\frac{E_{\Lambda}}
{M_{\Lambda}} +1)p_{\pi}^3} .
\eers
From the above we obtain $ g \approx 1.13$.  The  contribution of
the $\Sigma^*$  exchange using the propagator  $S_{\mu\nu}^1$ can
be written as
\ber
f_S & = & -0.127 +0.059z +0.089 z^2 , \nonumber\\
f_P & = & 0.119 -0.012z +0.054 z^2 . \
\eer
We observe that the  contribution  from the offshell terms can be
almost of the same  order as the  other  terms if $z \sim 1$.  If
$z$ is large  with $z > 1$, say  $z\sim 2 -3$, then the  offshell
term can even dominate the other terms.  However this is unlikely
as it would correspond to anomalously  large SU(3) breaking since
the magnitude  \cite{Nimai}  of the off--shell  coupling in the $
\pi N \Delta$ system is $< 1$.  In the heavy baryon framework Eq.
(20) indicates  that HBCHPT with baryons is slowly  converging as
in the case of the $SU(2)_L \times SU(2)_R$ HBCHPT describing the
pion-nucleon   system   \cite{DatPak}.  In  the  figure  we  plot
$\delta_S$ and $\delta_P$ from the $\Sigma^*$  contribution  as a
function of $z$.  Studies in the $\pi N \Delta$  system  obtain a
range  of  $z$ to be  between  -0.3  to  0.8  from  fit a to  the
pion--nucleon  data  \cite{Nimai}.  In the SU(3) limit we can use
this  range  for  our  calculations.  Assuming  reasonable  SU(3)
breaking  effects  we take the  range  of $z$ in our  calculation
between  -1 and 1.  From the  figure  we see that  $\delta_P$  is
always positive and $\delta_S$ is mostly negative for most values
of $z$ within  the range of -1 to 1.  The total  phase  shifts in
$\Lambda-\pi$  scattering is to a good  approximation  the sum of
the phase shifts from the $\Sigma$ and  $\Sigma^*$  exchange.  So
the net  $\delta_S$ and $\delta_P$ can have values  between $\sim
-1.3$ to $0.1$  degrees and  between  $\sim -3$ degrees to $-0.4$
degrees, respectively.

If we use the  propagator  $S_{\mu\nu}^2$  then we get $\delta_S=
$0.62 degrees and  $\delta_P=$  1.12 degrees from the  $\Sigma^*$
exchange.  This leads to total phase shifts of  $\delta_S=$  0.53
to 0.44  degrees  and  $\delta_P  =$ -0.8 to  -2.8  degrees 
independent
of the value of $z$.  The
inclusion  of the next low lying  negative  parity  $\Sigma$  and
$\Sigma^*$ resonances in our calculations is not expected to make
a significant contribution to the phase shifts \cite{ml}.

We now compare our results with those  obtained in a recent 
calculation of the phase shifts using $SU(3)_L \times SU(3)_R$  
chiral perturbation theory\cite{Kamal}.  We
 get very similar results as that of  \cite{Kamal}  if we neglect  the
off--shell   coupling  of  the   $\Sigma^*$   and  set  the  pion
mass to zero. If one expands the phase shifts in the parameter
$x=\frac{p_{\pi}}{M} \sim 0.14$ where $p_\pi$ is the the pion 3-momentum and 
$M$ the baryon mass then the S wave phase shift is generally suppressed
by a factor of $x$ compared to the P wave phase shift. While there 
might  be compensation of this suppression if in 
the expression for  the S wave phase shift
  $x$ is associated 
with a large coefficient but it is unlikely 
that the S wave phase shift can be much larger then the P wave phase shift.
Our results are consistent with this expectation. 
   
 In  summary  we have  calculated  the S and P phase  shifts  for
$\Lambda-\pi$  scattering  in  the  fully  relativistic  $SU(3)_L
\times SU(3)_R$  invariant  chiral  Lagrangian.  We also included
possible  off--shell   couplings  of  the  $\Sigma^*$  baryon  to
$\Lambda   \pi$.  Assuming   reasonable   SU(3)   breaking   this
off--shell  coupling  is taken to be of the same  order as in the
$\pi N \Delta$ system.  We find small phase shifts for both the S
and the P waves  which are of the same order as those  calculated
using $SU(2)_L \times SU(2)_R$ invariant chiral Lagrangian in the
heavy baryon limit.

{\bf Acknowledgment:}
 We thank Mahiko Suzuki and Salam Tawfiq for useful discussions. 
This  work  was  supported  in  part  by  National   Science  and
Engineering  Research  Council  of Canada  (A. Datta  and  Patrick
O'Donnell), Fermilab Theory Division's Summer Visitor Program (A. Datta)
 and DOE grant USA (S.  Pakvasa).

\subsection{Figure Caption}
\begin{itemize}

\item {\bf Figure}: S and P wave phase shifts $\delta_S$ 
and $\delta_P$ from $\Sigma^*$ exchange versus the offshell coupling  $z$.

 \end{itemize}

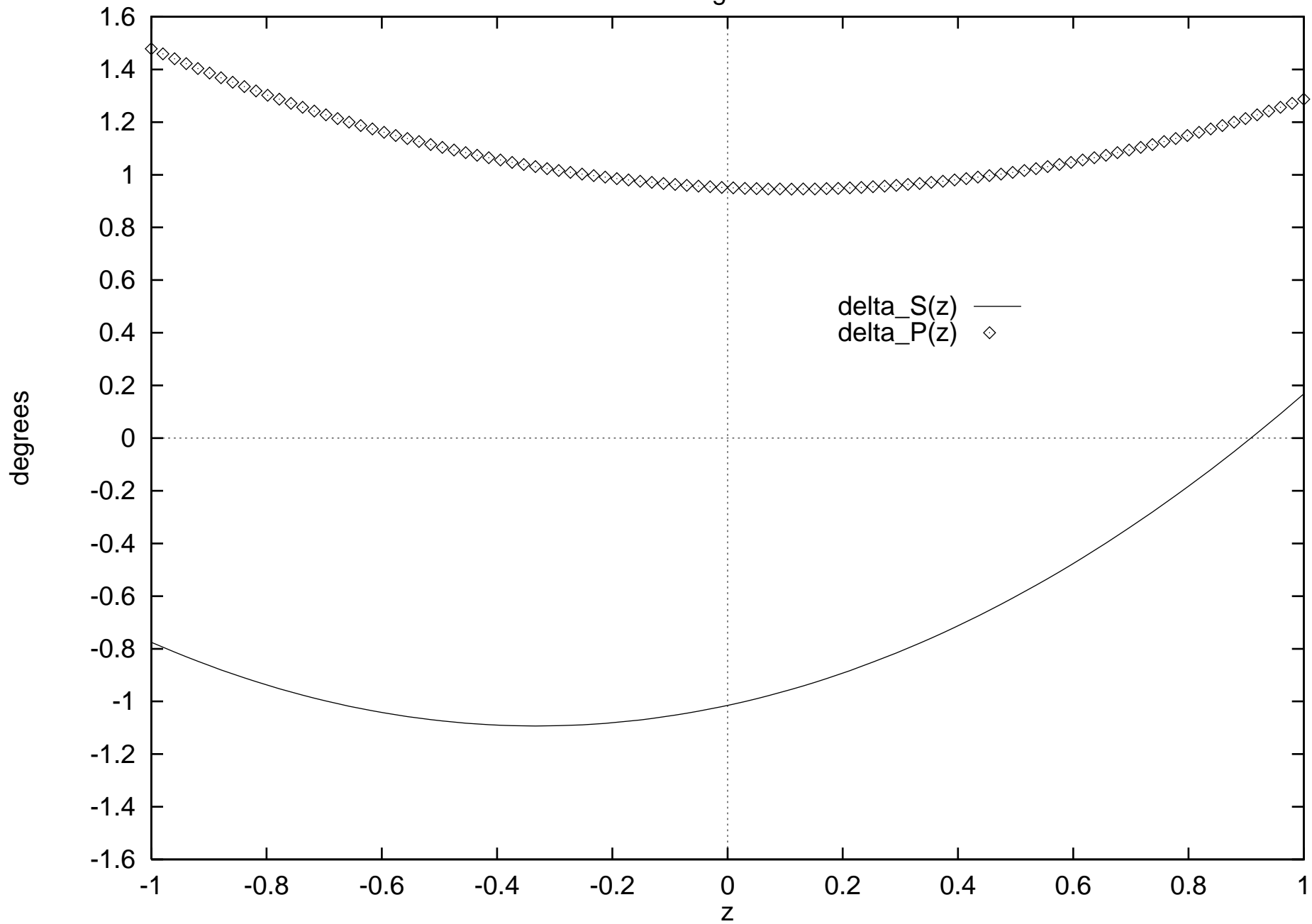
\end{document}